\documentclass[runningheads]{llncs}

\usepackage[square,comma,numbers,sort&compress,sectionbib]{natbib}
\usepackage[colorlinks,citecolor=blue,linkcolor=blue]{hyperref}

\usepackage{amsmath}
\usepackage{amssymb}
\usepackage{bbm}
\usepackage{booktabs}
\usepackage{multirow}
\usepackage{graphicx}
\usepackage{xcolor}
\usepackage[capitalize]{cleveref}
\usepackage{fancyhdr}

\fancypagestyle{specialfooter}{%
  \fancyhf{}
  
  \fancyfoot[C]{\vspace{1cm} \hspace*{-.2\textwidth}\parbox{1.4\textwidth}{\small This is the author’s version of the work. It is posted here for your personal use. The definitive version is published in: \emph{Proceedings of the 41th European Conference on Information Retrieval} (ECIR '19), April 14--18, 2019, Cologne, Germany}}
}

\makeatletter
\renewcommand\@biblabel[1]{#1.}
\makeatother

\newcommand{\up}{$^{\vartriangle}$}
\newcommand{\upup}{$^{\blacktriangle}$}
\newcommand{\down}{$^{\triangledown}$}
\newcommand{\downdown}{$^{\blacktriangledown}$}
\newcommand{\nosign}{$^{\circ}$}

\newcommand{\Cb}[1]{\colorbox{green!40}{#1}}

\begin{document}
\title{Identifying Unclear Questions in Community Question Answering Websites}
\author{Jan Trienes\inst{1} \and Krisztian Balog\inst{2}}
\authorrunning{J. Trienes and K. Balog}
\institute{University of Twente, Enschede, Netherlands \and
University of Stavanger, Stavanger, Norway\\
\email{jan.trienes@gmail.com},
\email{krisztian.balog@uis.no}}
\maketitle
\begin{abstract}
Thousands of complex natural language questions are submitted to community question answering websites on a daily basis, rendering them as one of the most important information sources these days.
However, oftentimes submitted questions are unclear and cannot be answered without further clarification questions by expert community members.
This study is the first to investigate the complex task of classifying a question as clear or unclear, i.e., if it requires further clarification.
We construct a novel dataset and propose a classification approach that is based on the notion of similar questions.
This approach is compared to state-of-the-art text classification baselines.
Our main finding is that the similar questions approach is a viable alternative that can be used as a stepping stone towards the development of supportive user interfaces for question formulation.
\end{abstract}
\thispagestyle{specialfooter}

\section{Introduction}
The emergence of community question answering (CQA) forums has transformed the way in which people search for information on the web.
As opposed to web search engines that require an information seeker to formulate their information need as a typically short keyword query, CQA systems allow users to ask questions in natural language, with an arbitrary level of detail and complexity.
Once a question has been asked, community members set out to provide an answer based on their knowledge and understanding of the question. Stack Overflow, Yahoo! Answers and Quora depict popular examples of such CQA websites.

\begin{figure}[t]
\centering
\fbox{\includegraphics[width=.95\textwidth]{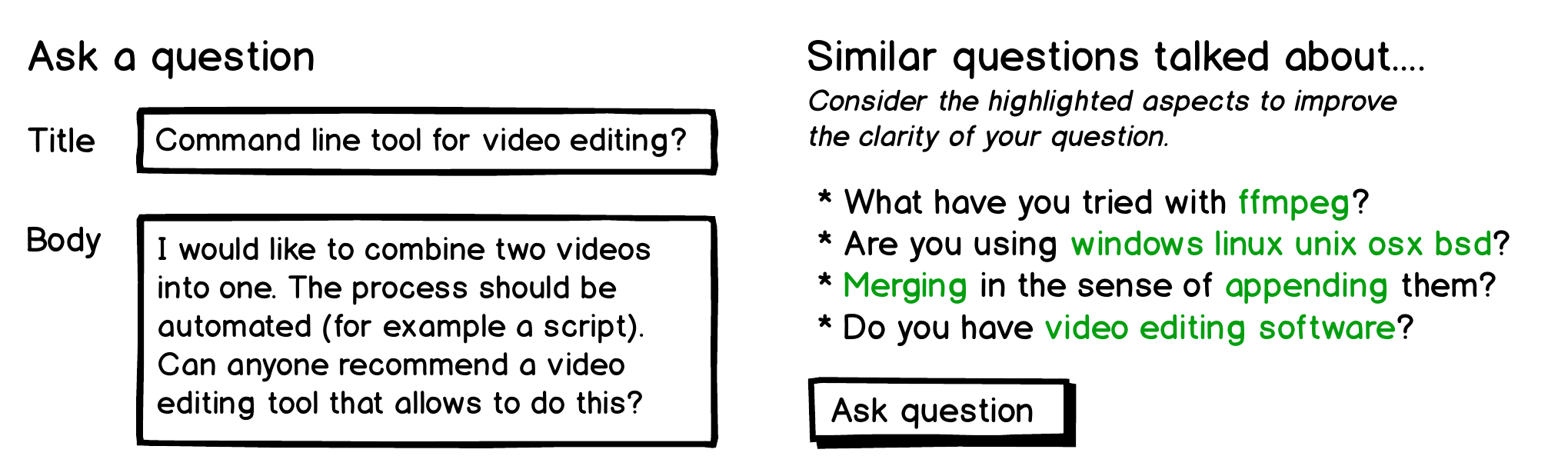}}
\caption{Envisioned question formulation interface. If a question is found to be unclear, a list of clarification questions (obtained from similar questions) is presented to encourage the user to include information that may be required to provide an answer.}
\label{fig:qac-ui}
\end{figure}

Despite their growing popularity and well established expert communities, increasing amounts of questions remain ignored and unanswered because they are too short, unclear, too specific, hard to follow or they fail to attract an expert member~\cite{Asaduzzaman:2013:AQU}.
To prevent such questions from being asked, the prediction of question quality has been extensively studied in the past~\cite{Ravi:2014:GQQ,Ponzanelli:2014:UCQ,Arora:2015:GBK}.
Increasing the question quality has a strong incentive as it directly affects answer quality~\cite{Li:2012:APQ}, which ultimately drives the
popularity and traffic of a CQA website.
However, such attempts ignore the fact that even a high-quality question may lack an important detail that requires clarification. 
On that note, previous work attempts to identify what aspects of a question requires editing.
While the need for editing can be reliably detected, the prediction of whether or not a question lacks important details has
been shown to be difficult~\cite{Yang:2014:ARQ}.
In order to support an information seeker in the formulation of her question and to increase question quality, we envision the following two-step system: (1) determine whether a question requires clarification (i.e., it is unclear), and (2) automatically generate and ask clarifying questions that elicit the missing information.
This paper addresses the first step. When successful, the automated identification of unclear questions is believed to have a strong impact on CQA websites, their efforts to increase question quality and the overall user experience; see~\cref{fig:qac-ui} for an illustration.

We phrase the unclear question detection as a supervised, binary classification problem and introduce the Similar Questions Model (SQM), which takes characteristics of similar questions into account.
This model is compared to state-of-the-art text classification baselines, including a bag-of-words model and a convolutional neural network.
Our experimental results show that this is a difficult task that can be solved to a limited extent using traditional text classification models.
SQM provides a sound and extendable framework that has both comparable performance and promising options for future extensions.
Specifically, the model can be used to find keyphrases for question clarification that may be utilized in a question formulation interface as shown in~\cref{fig:qac-ui}.
Experiments are conducted on a novel dataset including more than 6 million labeled Stack Exchange questions, which we release for future research on this task.\footnote{The dataset and sources can be found at \url{https://github.com/jantrienes/ecir2019-qac}}

\section{Related Work}
Previous work on question modeling of CQA forums can be roughly grouped into three categories: \textit{question quality prediction}, \textit{answerability prediction} and \textit{question review prediction}~\cite{Srba:2016:CSC}.
With respect to the prediction of question quality, user reputation has been found to be a good indicator~\cite{Tausczik:2011:PPQ}. Also, several machine learning techniques have been applied including topic and language models~\cite{Ravi:2014:GQQ,Arora:2015:GBK}.
However, there is no single objective definition of quality, as such a definition depends on the community standards of the given platform.
In this paper, we do not consider question quality itself, since a question may lack an important detail regardless of whether its perceived quality is high or low.
Question answerability has been studied by inspecting unanswered questions on Stack Overflow~\cite{Asaduzzaman:2013:AQU}. Lack of clarity and missing information is among the top five reasons for a question to remain unanswered.
Here, we do not consider other problems such as question duplication and too specific or off-topic questions~\cite{Correa:2014:CWC}.
Finally, question review prediction specifically attempts to identify questions that require future editing. Most notably,~\citet{Yang:2014:ARQ} determine if a question lacks a code example, context information or failed solution attempts based on its contents.
However, they disregard the task of predicting whether detail (e.g., a software version identifier) is missing and limit their experiments to the programming domain.

Clarification questions have been studied in the context of synchronous Q\&A dialog systems.~\citet{Kato:2013:CQS} analyzed how clarification requests influence overall dialog outcomes. In contrast to them, we consider asynchronous CQA systems.
With respect to asynchronous systems,~\citet{Braslavski:2017:YME} categorized clarification questions from two Stack Exchange domains.
They point out that the detection of unclear questions is a vital step towards a system that automatically generates clarification questions. To the best of our knowledge, we are the first study to address exactly this novel unclear question detection task.
Finally, our study builds on recent work by~\citet{Rao:2018:LAG}. We extend their dataset creation heuristic to obtain both clear and unclear questions.

\section{Unclear Question Detection}
The unclear question detection task can be seen as a binary classification problem. Given a dataset of $N$ questions, $Q = \{q_1, ..., q_N\}$, where each question belongs to either the \emph{clear} or \emph{unclear} class, predict the class label for a new (unseen) question $q$.
In this section, we propose a model that utilizes the characteristics of similar questions as classification features.
This model is compared to state-of-the-art text classification models described in~\cref{sec:methods}.

We define a question to be \emph{unclear} if it received a clarification question, and as \emph{clear} if an answer has been provided without such clarification requests.
This information is only utilized to obtain the ground truth labels.
Furthermore, it is to be emphasized that it is most useful to detect unclear questions during their creation-time in order to provide user feedback and prevent unclear questions from being asked (see the envisioned use case in~\cref{fig:qac-ui}).
Consequently, the classification method should not utilize input signals available only after question creation, such as upvotes, downvotes or conversations in form of comments.
Finally, we do not make any assumptions about the specific representation of a question as it depends on the CQA platform at hand.
The representation we employ for our experiments on Stack Exchange is given in~\cref{sec:dataset-creation}.

\begin{table}[t]
\centering
\caption{Example of a new unclear question (Left) and a similar existing question (Right). The left question fails to specify the operating system. The text in italics has been added in response to the shown comment.}
\label{tab:similar-questions-example}
\setlength{\tabcolsep}{0.5em}
\begin{tabular}{@{}lp{.37\textwidth}p{.42\textwidth}@{}}
\toprule
Field & New Question & Similar (Unclear) Question \\ \midrule
Title & Simplest XML editor & XML Editing/Viewing Software \\
Body & I need the simplest editor with utf8 support for editing xml files; It's for a non programmer (so no atom or the like), to edit existing files. Any suggestion? & What software is recommended for working with and editing large XML schemas? \emph{I'm looking for both Windows and Linux software (doesn't have to be cross platform, just want suggestions for both)} \\
Tags & xml, utf8, editors & windows, xml, linux \\
Comments & -- & What operating system?
\\ \bottomrule
\end{tabular}
\end{table}

\begin{figure}[t]
\centering
\fbox{\includegraphics[width=.98\textwidth]{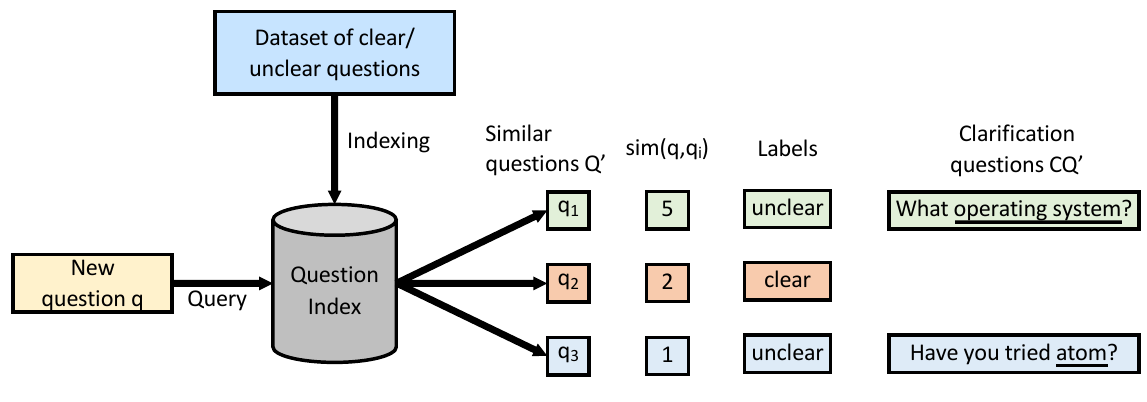}}
\caption{Illustration of Similar Questions Model. The underlined text of a clarification question indicates a keyphrase.}
\label{fig:simq-model}
\end{figure}

\subsection{Similar Questions Model}
The Similar Questions Model is centered around the idea that similar existing questions may provide useful indicators about the presence or absence of information.
For example, consider the two questions in~\cref{tab:similar-questions-example}. It can be observed that the existing question specifies additional information after a clarification question has been raised.
A classification system may extract keyphrases (e.g., \emph{operating system}) from the clarification question and check whether this information is present in the given question (see~\cref{fig:qac-ui} and~\cref{tab:simq-examples} for examples).
In other words, the system checks if a new question lacks information that was also missing from similar previous questions.
It has been shown that this general approach can be successfully employed to find and rank suitable clarification questions~\cite{Rao:2018:LAG}.\looseness=-1

\begin{table}[t]
\caption{Features employed by the Similar Questions Model. The example values in the last column are based on the scenario presented in~\cref{tab:similar-questions-example} and~\cref{fig:simq-model}.}
\label{tab:features}
\begin{tabular}{@{}lll@{}}
\toprule
\multicolumn{2}{l}{(i) Features based on $q$} & Ex. \\ \midrule
$Len(q)$ & Question length in the number of tokens: $|q|$ &  41\\
$ContainsPre(q)$ & Indicator if question contains preformatted elements & 0\\
$ContainsQuote(q)$ & Indicator if question contains a quote & 0\\
$ContainsQuest(q)$ & Indicator if question contains question mark ``?'' & 1\\
$Readability(q)$ & Coleman-Liau Index (CLI)~\cite{Coleman:1975:CRF} &  16.7\\ \midrule
\multicolumn{2}{l}{(ii) Features based on $Q'$} &  \\ \midrule
$SimSum(q, Q')$ & Sum of similarity scores: $\sum_{q' \in Q'} sim_{BM25}(q, q')$ &  8\\
$SimMax(q, Q')$ & Maximum similarity: $\max_{q' \in Q'} sim_{BM25}(q, q')$ &  5\\
$SimAvg(q, Q')$ & Average similarity: $\frac{1}{|Q'|} \sum_{q' \in Q'} sim_{BM25}(q, q')$ &  2.7\\
$LenSim(Q')^{\dag}$ & Number of similar questions retrieved: $|Q'|$ &  3\\
$LenUnclear(Q')^{\dag}$ & Number of similar questions that are unclear: $|Q'_{unclear}|$ &  2\\
$LenClear(Q')^{\dag}$ & Number of similar questions that are clear: $|Q'_{clear}|$ &  1\\
$Majority(Q')^{\dag}$ & Majority vote of labels in $Q'$ &  1\\
$Ratio(Q')^{\dag}$ & Ratio between clear/unclear questions: $|Q'_{clear}|/|Q'_{unclear}|$ &  0.5 \\
$Fraction(Q')^{\dag}$ & Proportion of clear questions among similar: $|Q'_{clear}|/|Q'|$ &  0.3\\ \midrule
\multicolumn{2}{l}{(iii) Features based on $CQ'^{\ddag}$} &  \\ \midrule
$CQGlobal(q, CQ')$ & Cosine similarity between all keyphrases in $CQ'$ and $q$ &  0.6\\
$CQIndividual(q, CQ')$ & Sum of cosine similarities between each keyphrase and $q$ &  1\\
$CQWeighted(q, CQ')$ & Like above, but weighted by $sim_{BM25}(q, q')$, see~\cref{eq:clarq-weighted} & 1 \\ \bottomrule \\
\multicolumn{3}{l}{$^{\dag}$\scriptsize{
These features are computed for the top-$k$ similar questions in $Q'$ where $k = \{10,20,50\}$.
}}\\
\multicolumn{3}{l}{$^{\ddag}$\scriptsize{%
$CQ'$ is obtained from the top $k=10$ similar questions in $Q'_{unclear}$.
}}
\end{tabular}
\end{table}

The Similar Questions Model can be formalized as follows. Given a new question $q$, we first seek a set of $k$ similar questions $Q' =\{q'_1, q'_2, ..., q'_k\}$ with their clear and unclear labels.
As per the definition of unclear that we employ, the subset of unclear questions $Q'_{unclear}$ has a set of $M$ corresponding clarification questions $CQ' =\{cq'_1, cq'_2, ..., cq'_M\}$.
Within this framework, we design a number of indicative features that are then used to train a classifier to discriminate between the two classes. An illustration of the model can be found in~\cref{fig:simq-model}.

\subsection{Features}
The features employed by the Similar Questions Model can be grouped into three classes: (i) features based on $q$ only, (ii) features based on the set of similar questions $Q'$ and (iii) features based on the set of clarification questions $CQ'$. See \cref{tab:features} for a summary.
We highlight the computation of the scoring features obtained from the set of clarification questions (group (iii) in~\cref{tab:features}).
For each clarification question in $CQ'$, one or more keyphrases are extracted. These keyphrases are the central objects of a clarification question and refer to an aspect of the original question that is unclear (see~\cref{tab:simq-examples} for examples).
Afterwards, we define $f(a) = (p_1, ..., p_i, ..., p_L)$ to represent a question or clarification question as a vector, where each element indicates the number of times a keyphrase $p_i$ occurs in $q$ and $cq'$, respectively.
Then, a \emph{question clarity score} is obtained by computing the cosine similarity between these vectors.
The scoring features differ in the way the keyphrase vectors are created.
The global model constructs a single vector consisting of all keyphrases present in $CQ'$, whereas the individual model computes the sum of the scores considering each $cq' \in CQ'$ separately.
For the individual weighted feature, the final score is given by:
\begin{equation}
CQWeighted(q, CQ') = \sum_{cq' \in CQ'} sim_{cos}(f(q), f(cq')) sim_{BM25}(q,q'),
\label{eq:clarq-weighted}
\end{equation}
where $sim_{cos}$ is the cosine similarity between the keyphrase vectors and $sim_{BM25}$ is the similarity between $q$ and $q'$.
This gives higher importance to keyphrases belonging to more similar questions.

\subsection{Learning Method}
We operationalize the Similar Questions Model in a variety of ways:
\begin{description}
	\item[SimQ Majority] We obtain a simple baseline that classifies $q$ according to the most common label of the similar questions in $Q'$.
	\item[SimQ Threshold] We test the scoring features in group (iii) using a threshold classifier where a threshold $\gamma$ is learned on a held-out dataset. The label is then obtained as follows:
	\begin{equation*}
\hat{y} =
	\begin{cases}
      0, & \text{if}\ feat(q, CQ') \geq \gamma \\
      1, & \text{otherwise},
    \end{cases}
\end{equation*}
where $feat(q, CQ')$ is the value of the corresponding feature, 0 refers the clear class and 1 refers to the unclear class.
	\item[SimQ ML] All features of the Similar Questions Model are combined and provided as input data to a machine learning classifier.
\end{description}

\section{Experimental Setup}
This section describes our experimental setup including the dataset and methods.

\subsection{Dataset Creation}
\label{sec:dataset-creation}
The Stack Exchange CQA platform depicts a suitable data source for our experiments. It is a network of specialized communities with topics varying from programming to Unix administration, mathematics and cooking.
A frequent data dump is published consisting of all questions, answers and comments submitted to the site. For any post, a time-stamped revision history is included.
We use this dump\footnote{Available at \url{https://archive.org/details/stackexchange}} to create a labeled dataset consisting of clear and unclear questions.

To obtain unclear questions, we apply a heuristic that has been used in previous research to find clarification questions~\cite{Braslavski:2017:YME, Rao:2018:LAG}.
A question is considered to be unclear when there is a comment by a different user than the original asker and that comment contains a sentence ending with a question mark.
This heuristic is not perfect as it will inevitably miss clarification requests not formulated as a question (e.g., \emph{``Please post your code.''}), while it retains rhetorical questions (e.g., \emph{``Is this a real question?''}).
We only keep those questions where the original asker has provided a clarification in form of a comment or question edit.

In order to gather clear questions, we extend the described heuristic as follows. A question is considered to be clear if it has neither edits, nor comments, but it has an accepted answer.
An answer can be manually accepted by the question asker if they consider it to adequately answer their question. Again, this heuristic may introduce noise: an answer can make certain assumptions that would have ideally been asked as a clarification question instead of included in the answer itself (e.g., \textit{``Provided you are on system X, the solution is Y''}).

We apply this heuristic to five Stack Exchange communities, each of a different size and with a different domain.
The communities considered are Stack Overflow, Ask Ubuntu, Cross Validated, Unix \& Linux and Super User, thus covering a broad range of topics.
\cref{tab:dataset-statistics} summarizes the statistics of each dataset.
The text has been preprocessed by combining the question title, body and tags into a single field, replacing URLs with a special token, converting every character to lower-case and removing special characters except for punctuation.
Token boundaries are denoted by the remaining punctuation and whitespace. Furthermore, a minimum term-document frequency of 3 is imposed to prevent overfitting.

\begin{table}[t]
\centering
\caption{Dataset statistics and class distribution. $N$ is the number of samples, $L$ the median sample length in tokens, and $|V|$ the vocabulary length after tokenization. $|V^*|$ is the vocabulary length with an imposed minimum term-document frequency of 3.}
\label{tab:dataset-statistics}
\setlength{\tabcolsep}{0.5em}
\begin{tabular}{@{}lcccccc@{}}
\toprule
Community & $N$ & $L$ & $|V|$ & $|V^*|$ & Clear & Unclear \\ \midrule
Stack Overflow & 5,859,667 & 159 & 8,939,498 & 1,319,587 & 35\% & 65\% \\
Super User & 121,998 & 121 & 206,249 & 45,432 & 33\% & 67\% \\
Ask Ubuntu & 77,712 & 114 & 188,476 & 40,309 & 27\% & 73\% \\
Unix \& Linux & 44,936 & 133 & 162,805 & 31,852 & 27\% & 73\% \\
Cross Validated & 38,488 & 157 & 130,691 & 24,229 & 18\% & 82\% \\ \bottomrule
\end{tabular}
\end{table}

\subsection{System Components}
\label{sec:methods}

\subsubsection{Obtaining Similar Questions}
A general purpose search engine, Elasticsearch, is used with the BM25 retrieval model in order to obtain similar questions.
The retrieval score is used as $sim_{BM25}(q,q')$ during feature computation.
We only index the training set of each community but retrieve similar questions for the entire dataset.
Queries are constructed by combining the title and tags of a question. These queries are generally short (averaging 13 tokens).\footnote{We experimented with longer queries that include 100 question body tokens. While computationally more expensive, model performance remained largely unaffected.}
To ensure efficient querying, we remove stopwords from all queries.
Finally, BM25 parameters are set to common defaults ($k_1=1.2$, $b=0.75$)~\cite{Manning:2008:IIR}.

\subsubsection{Extracting Keyphrases}
Keyphrases are extracted from clarification questions using the RAKE algorithm~\cite{Rose:2010:AKE}, which is an efficient way to find noun phrases.
This algorithm has been used in a similar setting where CQA comments should be matched to related questions~\cite{Nandi:2017:EMF}.
We tokenize the keyphrases and consider each token individually.

\subsubsection{Similar Questions Classifier}
Besides applying a threshold-based classifier on a selected set of features presented in~\cref{tab:features}, all features are combined to train a logistic regression classifier with L2 regularization (referred to as SimQ ML).
The regularization strength is set to $C=1$ which has been found to work well for all communities.
All features are standardized by removing the mean and scaling to unit variance.

\subsection{Baseline Models}
The Similar Questions Model is compared with a number of baselines and state-of-the-art text classification approaches:
\begin{itemize}
	\item Random: produce predictions uniformly at random.
	\item Majority: always predict the majority class (here: \emph{unclear}).
	\item Bag-of-words logistic regression (BoW LR).
	\item Convolutional neural network (CNN)~\cite{Kim:2014:CNN}.
\end{itemize}
Within the BoW LR model, a question is represented as a vector of TF-IDF weighted n-gram frequencies. Intuitively, this approach captures question clarity on a phrase and topic level.
We report model performances for unigrams ($n=1$) and unigrams combined with phrases of length up to $n=3$.
Using 5-fold cross-validation on the training data, we find that an L2 regularization strength of $C=1$ works best for all communities.
With respect to the CNN model, we use the static architecture variant presented in~\cite{Kim:2014:CNN} consisting of a single convolutional layer, followed by a fully connected layer with dropout.
Model hyperparameters (number of filters, their size, learning rate and dropout) are optimized per community using a development set.\footnote{Optimal CNN parameter settings can be found in the online appendix of this paper.}
The network is trained with the Adam optimizer~\cite{Kingma:2015:AMS}, a mini-batch size of 64 and early stopping. We train 300-dimensional word embeddings for each community using word2vec~\cite{Mikolov:2013:DRW} and limit a question to its first 400 tokens (with optional padding).
Out-of-vocabulary words are replaced by a special token.
There are several other possible neural architectures, but an exploration of those is outside the scope of this paper.

\subsection{Evaluation}
As the data is imbalanced, we evaluate according to the F1 score of the unclear (positive) class and the ROC AUC score.
We argue that it is most important to optimize these metrics based on the envisioned use case. When the classification outcome is used as a quality guard in a user interface, it is less sever to consider a supposedly clear question as unclear as opposed to entirely missing an unclear question.
We randomly divide the data for each community into 80\% training and 20\% testing splits. Of the training set, we use 20\% of the instances for hyperparameter tuning and optimize for ROC AUC.
We experimented with several class balancing methods, but the classification models were not impacted negatively by the (slight) imbalance.
Statistical significance is tested using an approximate randomization test.
We mark improvements with \up ($p<0.05$) or \upup ($p<0.01$), deteriorations with \down ($p<0.05$) or \downdown ($p<0.01$), and no significance by \nosign.\looseness=-1

\section{Results and Analysis}
This section presents and discusses our experimental results.

\begin{table}[t]
\centering
\caption{Results for unclear question detection. The metrics are summarized over the five datasets using both micro-averaging and macro-averaging. F1, precision and recall are reported for the unclear class. Best scores for each metric are in boldface.}
\label{tab:evaluation-summary}
\begin{tabular}{@{}p{3cm}llllllll@{}}
\toprule
& \multicolumn{4}{c}{Micro-average} & \multicolumn{4}{c}{Macro-average} \\
\cmidrule(lr){2-5}
\cmidrule(lr){6-9}
Method & Acc. & F1 & Prec. & Rec. & Acc. & F1 & Prec. & Rec. \\ \midrule
Random & 0.499 & 0.564 & 0.649 & 0.499 & 0.497 & 0.586 & 0.714 & 0.499 \\
Majority & 0.649 & 0.787 & 0.649 & \textbf{1.000} & 0.719 & 0.835 & 0.719 & \textbf{1.000} \\
BoW LR ($n=1$) & 0.687 & 0.786 & 0.706 & 0.886 & 0.736 & 0.833 & \textbf{0.752} & 0.933 \\
BoW LR ($n=3$) & \textbf{0.699} & 0.791 & \textbf{0.720} & 0.877 & \textbf{0.741} & \textbf{0.837} & \textbf{0.752} & 0.944 \\
CNN & \textbf{0.699} & \textbf{0.794} & 0.715 & 0.893 & 0.739 & 0.836 & 0.749 & 0.947 \\ \midrule
SimQ Models \\ \midrule
SimQ Majority & 0.566 & 0.673 & 0.659 & 0.688 & 0.676 & 0.780 & 0.739 & 0.826 \\
CQ Global & 0.594 & 0.727 & 0.645 & 0.833 & 0.626 & 0.753 & 0.713 & 0.803 \\
CQ Individual & 0.586 & 0.721 & 0.640 & 0.824 & 0.632 & 0.761 & 0.710 & 0.824 \\
CQ Weighted & 0.604 & 0.737 & 0.648 & 0.855 & 0.642 & 0.770 & 0.716 & 0.838 \\
SimQ ML & 0.673 & 0.781 & 0.690 & 0.902 & 0.728 & 0.833 & 0.736 & 0.960 \\ \bottomrule\end{tabular}
\end{table}

\subsection{Results}
The traditional BoW LR model provides a strong baseline across all communities that outperforms both the random and majority baselines (see~\cref{tab:results-by-community}).
The generic CNN architecture proposed in~\cite{Kim:2014:CNN} does not provide any significant improvements over the BoW LR model.
This suggests that a more task-specific architecture may be needed to capture the underlying problem.

We make several observations with respect to the Similar Questions Model. First, a majority vote among the labels of the top $k=10$ similar questions (SimQ Majority) consistently provides a significant improvement over the random baseline for all datasets (see~\cref{tab:results-by-community}).
This simplistic model shows that the underlying concept of the Similar Questions Model is promising.
Second, the scoring features that take clarification questions into consideration do not work well in isolation (see models prefixed with CQ in~\cref{tab:evaluation-summary} and~\cref{tab:results-by-community}).
The assumption that one can test for the presence of keyphrases without considering spelling variations or synonyms seems too strong.
For example, the phrase \emph{``operating system''} does not match sentences such as \emph{``my OS is X''} and thus results in false positives.
Finally, the SimQ ML model outperforms both the random and majority baselines, and has comparable performance with the BoW LR model.
It is to be emphasized that the SimQ ML model, in addition to classifying a question as clear or unclear, generates several valuable hints about the aspects of a question that may be unclear or missing (see demonstration in~\cref{tab:simq-examples}).
This information is essential when realizing the envisioned user interface presented in~\cref{fig:qac-ui}, and cannot be deducted from the BoW LR or CNN models.

\begin{table}[t]
  \centering
  \caption{Model performance for a selected set of communities. F1 scores are reported for the unclear class. Significance for model in line $i > 1$ is tested against line $i - 1$. Additionally, significance of each SimQ model is tested against the BoW LR ($n=3$) model (second marker).}
  \resizebox{\textwidth}{!}{%
  \begin{tabular}{*{10}{l}}
    \toprule
    & \multicolumn{3}{c}{Cross Validated} & \multicolumn{3}{c}{Super User} & \multicolumn{3}{c}{Stack Overflow} \\
    \cmidrule(lr){2-4}
    \cmidrule(lr){5-7}
    \cmidrule(lr){8-10}
    Method & Acc. & AUC & F1 & Acc. & AUC & F1 & Acc. & AUC & F1 \\
    \midrule
Random & 0.493 & 0.500 & 0.618 & 0.502 & 0.500 & 0.575 & 0.499 & 0.500 & 0.563\\
Majority & 0.818\upup & 0.500\nosign & \textbf{0.900}\upup & 0.669\upup & 0.500\nosign & 0.802\upup & 0.646\upup & 0.500\nosign & 0.785\upup\\
BoW LR ($n=1$) & \textbf{0.819}\nosign & 0.647\upup & \textbf{0.900}\nosign & 0.702\upup & 0.720\upup & 0.798\nosign & 0.685\upup & 0.693\upup & 0.784\downdown\\
BoW LR ($n=3$) & 0.818\nosign & \textbf{0.659}\upup & \textbf{0.900}\nosign & \textbf{0.709}\upup & \textbf{0.731}\nosign & \textbf{0.807}\upup & \textbf{0.697}\upup & 0.718\upup & 0.788\upup\\
CNN & 0.817\nosign & 0.626\nosign & 0.899\nosign & 0.704\down & 0.715\nosign & 0.803\downdown & \textbf{0.697}\nosign & \textbf{0.720}\upup & \textbf{0.792}\upup\\
\midrule SimQ Models \\ \midrule
SimQ Majority & 0.796\downdown & 0.584\downdown & 0.883\downdown & 0.639\downdown & 0.616\downdown & 0.738\downdown & 0.561\downdown & 0.515\downdown & 0.667\downdown\\
CQ Global & 0.718\downdown\downdown & 0.515\downdown\nosign & 0.830\downdown\downdown & 0.598\downdown\downdown & 0.536\downdown\downdown & 0.733\nosign\downdown & 0.592\upup\downdown & 0.520\upup\downdown & 0.725\upup\downdown\\
CQ Individual & 0.713\nosign\downdown & 0.513\nosign\nosign & 0.827\nosign\downdown & 0.591\downdown\downdown & 0.549\upup\downdown & 0.728\downdown\downdown & 0.584\downdown\downdown & 0.528\upup\downdown & 0.719\downdown\downdown\\
CQ Weighted & 0.696\downdown\downdown & 0.496\downdown\nosign & 0.812\downdown\downdown & 0.602\upup\downdown & 0.534\downdown\downdown & 0.739\upup\downdown & 0.603\upup\downdown & 0.503\downdown\downdown & 0.736\upup\downdown\\
SimQ ML & \textbf{0.819}\upup\nosign & 0.631\nosign\nosign & \textbf{0.900}\upup\nosign & 0.687\upup\downdown & 0.671\upup\downdown & 0.798\upup\downdown & 0.670\upup\downdown & 0.666\upup\downdown & 0.779\upup\downdown\\
\bottomrule
  \end{tabular}
  }
  \label{tab:results-by-community}
\end{table}

\subsection{Feature Analysis}
To gain further insights about the performance of the Similar Questions Model, we analyze the features and their predictive power.
Features considering the stylistic properties of a question itself such as the length, readability and whether or not the question contains a question mark, are among the top scoring features (see~\cref{fig:feature-importance}).
Other important features include the distribution of labels among the similar questions and their retrieval scores ($LenUnclear$, $LenClear$, $SimSum$, $Fraction$).
With respect to the bag-of-words classifier, we observe that certain question topics have attracted more unclear questions. For example, a question about \emph{windows 10} is more likely to be unclear than a question about \emph{emacs}. Interestingly, also stylistic features are captured (e.g., a ``?'' token and the special URL token). Finally, this model reveals characteristics of well-written, clear questions. For example, if a user articulates their problem in the form of \emph{``difference between X and Y,''} such a question is more likely to belong to the clear class.
This suggests that it may be beneficial to include phrase-level features in the Similar Questions Model to improve performance.

\begin{table}[t]
\centering
\caption{Subsets of learned coefficients for the Similar Questions Model (Left) and BoW LR classifier (Right). Both are trained on the Super User community. Positive numbers are indicative for the unclear class and negative values for the clear class.}
\label{fig:feature-importance}
\begin{tabular}{@{}lp{4cm}ll@{}}
\toprule
\multicolumn{2}{c}{SimQ ML} & \multicolumn{2}{c}{BoW LR ($n=3$)} \\
\cmidrule(lr){1-2}
\cmidrule(lr){3-4}
Coef. & Feature & Coef. & Feature \\ \midrule
+0.269 & $Len(q)$ & +4.321 & windows 10 \\
+0.166 & $CQIndividual(q, CQ')$ & +3.956 & \textless URL\textgreater \\
+0.146 & $LenUnclear(Q')\, (k=50)$ & +3.229 & problem \\
+0.120 & $LenSim(Q')\, (k=20)$ & +2.413 & nothing \\
+0.094 & $SimSum(q, Q')$ & +2.150 & help me \\
+0.081 & $Readability(q)$ & +1.760 & unable to \\
-0.091 & $ContainsPre(q)$ & -1.488 & documentation \\
-0.114 & $CQIndividual(q, CQ')$ & -1.742 & difference between \\
-0.150 & $LenClear(Q')\, (k=50)$ & -1.822 & can i \\
-0.150 & $Fraction(Q')\, (k=50)$ & -1.841 & emacs \\
-0.171 & $ContainsQuest(q)$ & -3.026 & vista \\
-0.196 & $SimMax(q,Q')$ & -5.306 & ? \\ \bottomrule
\end{tabular}
\end{table}

\subsection{Error Analysis and Limitations}
The feature analysis above reveals a problem which is common to both the Similar Questions Model and the traditional BoW LR model.
Both models suffer from a topic bias. For example, a question about \emph{emacs} is more likely to be classified as clear because the majority of \emph{emacs} questions are clear.
Furthermore, stylistic features can be misleading. Consider a post on Stack Overflow that contains an error message. This post does not require an explicit use of a question mark as the implied question most likely is \emph{``How can this error message be fixed?''.}
It is conceivable to design features that take such issues into account.

A potential limitation of the Similar Questions Model is its reliance on the existence of similar questions within a CQA website. It is unclear how the model would perform in the absence of such questions.
It would make an interesting experiment to process a CQA dataset in chronological order, and measure how the model's effectiveness changes as more similar questions become available over time.
However, we leave the exploration of this idea to future work.

\begin{table}[t]
\caption{Example questions and their clarification questions retrieved by Similar Questions Model. The numbers in parenthesis indicate retrieval score $sim_{BM25}(p, q_i)$. Highlighted text corresponds to the keyphrases extracted by RAKE.}
\label{tab:simq-examples}
\begin{tabular}{@{}lp{.9\textwidth}@{}}
\toprule
Field & Text \\ \midrule
Title & Laptop randomly going in hibernate \\
Body & I have an Asus ROG G751JT laptop, and a few days ago my battery has died. The problem that I am encountering is that my laptop randomly goes to sleep after a few minutes of use even when plugged in [...].
\\
ClarQ & (20.01) Does this \Cb{happen} if you \Cb{boot} instead from an \Cb{ubuntu liveusb}?\\
& (17.92) Did you \Cb{enable} allow \Cb{wake timers} in \Cb{power options sleep}?\\
& (16.88) Can you \Cb{pop} the \Cb{battery} out of the \Cb{mouse}? \\
& (16.64) Which \Cb{OS} are you using? \\
& (16.02) Have you \Cb{scanned} your \Cb{system} for \Cb{malwares}?\\\midrule
Title & Does ZFS make sense as local storage? \\
Body & I was reading about ZFS and for a moment thought of using it in my computer, but than reading about its memory requirements I thought twice. Does it make sense to use ZFS as local or only for servers used as storage?\\
ClarQ & (36.11) What's wrong with more \Cb{redundancy}? \\
& (31.41) What \Cb{kind} of \Cb{data} are you trying to \Cb{protect}? \\
& (30.77) How are you \Cb{planning} on doing \Cb{backups} and or \Cb{disaster recovery}? \\
& (29.70) Is \Cb{SSD large} enough? \\\bottomrule
\end{tabular}
\end{table}

\section{Conclusion}
The paper represents the first study on the challenging task of detecting unclear questions on CQA websites.
We have constructed a novel dataset and proposed a classification method that takes the characteristics of similar questions into account.
This approach encodes the intuition that question aspects which have been missing or found to be unclear in previous questions, may also be unclear in a given new question.
We have performed a comparison against traditional text classification methods.
Our main finding is that the Similar Questions Model provides a viable alternative to these models, with the added benefit of generating cues as to why a question may be unclear; information that is hard to extract form traditional methods but that is crucial for supportive question formulation interfaces.

Future work on this task may combine traditional text classification approaches with the Similar Questions Model to unify the benefits of both.
Furthermore, one may start integrating the outputs of the Similar Questions Model into a clarification question generation system, which at a later stage is embedded in the user interface of a CQA site. As an intermediate step, it would be important to evaluate the usefulness of the generated cues as to why a question is unclear.
Finally, the work by~\citet{Rao:2018:LAG} provides a natural extension, by ranking the generated clarification questions in terms of their expected utility.

\subsubsection{Acknowledgments.} We would like to thank Dolf Trieschnigg and Djoerd Hiemstra for their insightful comments on this paper. This work was partially funded by the University of Twente Tech4People Datagrant project.

\bibliographystyle{abbrvnat}
\bibliography{bibliography}

\end{document}